\documentclass{sigchi}
\usepackage{balance}       
\usepackage{graphics}      
\usepackage[T1]{fontenc}   
\usepackage{txfonts}
\usepackage{mathptmx}
\usepackage{hyperref}
\usepackage{color}
\usepackage{booktabs}
\usepackage{textcomp}

\usepackage{microtype}        
\usepackage{ccicons}          

\usepackage{todonotes}

\usepackage{pstricks}
\usepackage{colortbl}
\usepackage{xifthen}
\graphicspath{ {images/} }

\usepackage{enumitem}
\setlist{nolistsep,leftmargin=*}

\def\plaintitle{SIGCHI Conference Proceedings Format}

\def\emptyauthor{}
\def\plainkeywords{Authors' choice; of terms; separated; by
  semicolons; include commas, within terms only; required.}

\makeatletter
\def\url@leostyle{%
  \@ifundefined{selectfont}{
    \def\UrlFont{\sf}
  }{
    \def\UrlFont{\small\bf\ttfamily}
  }}
\makeatother
\def\pprw{8.5in}
\def\pprh{11in}

\definecolor{linkColor}{RGB}{6,125,233}
\hypersetup{%
  pdftitle={\plaintitle},
  pdfauthor={\emptyauthor},
  pdfkeywords={\plainkeywords},
  pdfdisplaydoctitle=true, 
  bookmarksnumbered,
  pdfstartview={FitH},
  colorlinks,
  citecolor=black,
  filecolor=black,
  linkcolor=black,
  urlcolor=linkColor,
  breaklinks=true,
  hypertexnames=false
}

\newcommand{\blind}[2][]{%
  \ifthenelse{\isempty{#1}}{[{\em anonymized for double-blind review}]}{#2}}

\newcommand{\hintmcq}{Hint\_MCQ}

\newcommand{\hintstring}{Hint\_String}
\newcommand{\p}[1]{#1\ensuremath{'}}

\newtheorem{example}{Example}[section]
\newcommand{\parseit}{\textit{ParseIT}}
\newcommand{\cmt}[1]{} 

\newcommand{\afterfig}{}
\begin{document}
    
\doi{}
\isbn{}
    
    
    \title{ParseIT: A Question-Answer based Tool to Learn Parsing Techniques}
    \numberofauthors{2}
    \author{
        \alignauthor
        Amey Karkare\\
        \affaddr{Indian Institute of Technology Kanpur}\\
        \affaddr{Kanpur, UP, India}\\
        \email{karkare@cse.iitk.ac.in}
        \alignauthor
         Nimisha Agarwal\\
        \affaddr{Indian Institute of Technology Kanpur}\\
        \affaddr{Kanpur, UP, India}\\
        \email{nimisha@cse.iitk.ac.in}
   }
    \maketitle
    
    \begin{abstract}
{\em Parsing} (also called {\em syntax analysis}) techniques
cover a substantial portion of any undergraduate Compiler Design
course.  We present \parseit, a tool to help students understand
the parsing techniques through question-answering.
\parseit\ automates the generation of tutorial questions based on
the Context Free Grammar provided by the student and generates
feedback for the student solutions.  The tool generates
multiple-choice questions (MCQs) and fill in the blank type
questions, and evaluates students' attempts. It provides hints
for incorrect attempts, again in terms of MCQs.  The hints
questions are generated for any correct choice that is missed or
any incorrect choice that is selected.  Another interesting form
of hint generated is an input string that helps the students
identify incorrectly filled cells of a parsing table. We also
present results of a user study conducted to measure the
effectiveness of \parseit.
\end{abstract}

    
    
    \section{Introduction}
Compiler  design  is an  important  subject  in the  computer  science
curriculum  for  undergraduates~\cite{cs2013}.  Compilers  are  one  of the  success
stories of  Computer Science,  where sound theoretical  concepts (e.g.
Automata, Grammars, Graph Theory, Lattice  Theory etc.)  are backed by
practical implementations (Lexical analyzers, Parsers, Code Optimizers
etc.)  to solve the real  world problem of fast and resource-efficient
compilation.             Most             existing            compiler
courses~\cite{asu,cs143,coursera.compilers,cs304}      divide      the
curriculum   into    modules   corresponding   to   the    phases   of
compilation. Instructors discuss the theory in lectures while students
typically work on a semester-long  project implementing a compiler for
some small language.

In a  typical course,  about 15\%-22\%  of the total  time is  spent on
syntax          analysis           phase (also called parsing
techniques, see~Table~\ref{tab:typicalCompiler}). A number  of concepts are
introduced  to explain  the internals  of parsers,  for example  first
sets, follow sets,  item set, goto and closure sets,  parse tables and
the   parsing   algorithms~\cite{asu},    making   the   understanding
difficult. While parser  generators (YACC and its  variants) allow the
students  to  experiment with  grammars,  the  working of  the  parser
generated by the tools  is still opaque\footnote{The generated parsers
  do produce debugging information when used with appropriate options,
  but this is of little didactical value as  one needs to
  know the parsing algorithms to understand it.}.

\begin{table}\centering
  \caption{Time Spent on Teaching Parsing.}\label{tab:typicalCompiler}
\begin{tabular}{|@{\ }p{13mm}@{}|@{\ }p{20mm}@{\ }|@{}r@{\ }|@{\ }r@{\ }|@{}r@{\ }|} \hline
  {\bf Institute} & {\bf Course Name} & \multicolumn{2}{@{}c|}{\bf \#/Duration of Lectures} & {\%} \\ \cline{3-4}
                    & & Parsing & Total & \\ \hline 
Stanford            & Intro. to \mbox{Compilers~\cite{cs143}}& 4 & 18 & 22\%\\ \hline
IIT Kanpur
& Compiler \mbox{Design~\cite{cs335}}
& 6 & 35 & 17\%\\ \hline
Coursera            & Compilers~\cite{coursera.compilers} & 4 modules & 18 modules & 22\% \\
                    &                                     & ($\equiv$4 hours) & ($\equiv$19 hours) & (21\%)\\ \hline
Saylor              & Compilers~\cite{cs304} & 28 hours & 146  hours & 19\% \\ \hline
\end{tabular}  
\afterfig
\end{table}

Recent development  in technologies  has enabled  institutions to
offer   courses    to   large    number   of    students.   These
massive-open-online  courses (MOOCs)~\cite{coursera,  edx, nptel}
digitize the contents of the  topics (lecture videos, notes etc),
and  allow  students  to  access  the  contents  beyond  physical
boundaries of classrooms.  The increase in number of students has
added  challenges for  the instructor  for the  tutoring aspects,
such as  the creation  of new  problems for  assignments, solving
these  problems,  grading,  and  helping the  students  master  a
concept  through  hands-on exercises.   These
challenges  have   prompted  researchers  to   develop  automated
tutoring systems to help the student to explore a course based on
his     skills     and    learning     speed~\cite{alur.automata,
  alur.automata2, autograder, mernik.lisa, sumit.cacm14}.

In  this paper,  we  present  \parseit, a  tool  for teaching  parsing
techniques.   \parseit\  helps  students  to  understand  the  parsing
concepts through automatically generated  problems and hints. Problems
are generated  based on a Context  Free Grammar (CFG) given  as input.
The  tool evaluates  the solutions  attempted  by the  user for  these
problems. Upon evaluation, if the  solutions provided by the users are
incorrect, it generates hint questions.  The problems generated by the
tool follow a general Multiple  Choice Question (MCQ) pattern, where a
user is given a  problem with a set of possible choices,  1 or more of
which  are correct.  The  incorrect  solutions are  the  ones where  a
correct option  is not chosen,  or an  incorrect option is  chosen, or
both.  The hints are generated  in the forms of (simplified) questions
to  direct  student  toward  the correct  solution.   Hint  generation
procedures involve different  types of algorithms, of  which the input
string generation algorithm is notable.   For an incorrect parse table
provided by the user, this algorithm  enables the creation of an input
string that distinguishes a successful parse from an unsuccessful one.

We  describe some  of  the  systems developed  by  other for  teaching
compiler  concepts  in  Sec.~\ref{sec:relwork}.  The  tool  itself  is
described   in   Sec.~\ref{sec:parseit}.   Input   string   generation
algorithms for  LL and LR  parsers are given  in Sec.~\ref{sec:input}.
We present  a summary of  the user study in  Sec.~\ref{sec:study}, and
conclude in Sec.~\ref{sec:concl}.

    \section{Related Work}\label{sec:relwork}
Several efforts exist to automate  teaching phases of compilers and to
help    developing     a    compiler    as    a     course    project.
LISA~\cite{mernik.lisa}  helps  students   learn  compiler  technology
through animations  and visualizations.   The tool uses  animations to
explain  the  working  of  3  phases  of  compilers,  namely,  lexical
analysis, syntax analysis, and semantic analysis.  Lexical analysis is
taught using animations in DFAs.   For syntax analysis, animations are
shown for the construction of  syntax trees and for semantic analysis,
animations are  shown for  the node  visits of  the semantic  tree and
evaluation of  attributes.  Students understand the  working of phases
by modifying the specification and observing the corresponding changes
in the animation.

Lorenzo et.   al.~\cite{Lorenzo} present a system  for test-case based
automated  evaluation of  compiler projects.   Test cases  (inputs and
corresponding desired outputs) designed by the instructor are given as
input to students' compilers.  The  tool then assesses the compiler in
three  distinct steps--compilation,  execution,  and correction.   The
system automatically generates different  reports (for instructors and
students) by analyzing the logs generated at each of these steps.

Demaille   et.     al.~\cite{demaille.05.iticse,   demaille.08.iticse}
introduce  several   tools  to   improve  the  teaching   of  compiler
construction projects  and make  it relevant  to the  core curriculum.
They made  changes to  Bison~\cite{bison} to provide  detailed textual
and  graphical descriptions  of the  LALR automata,  allow the  use of
named  symbols  in  actions  (instead  of \$1,  \$2,  etc.),  and  use
Generalized LR  (GLR) as  backend. Waite~\cite{Waite:2006}  proposed 3
strategies  for teaching  compilers--software project,  application of
theory  and support  for communicating  with a computer.   Various other
tools are  also available to  teach different phases of  compiler like
understanding code  generation~\cite{Sondag:2010},  and understanding
symbol tables through animations~\cite{Urquiza-Fuentes:2011}.

Our work  is different in that  we use question-answering as  means to
explain the  working of parsing  technology and to guide  the students
towards the construction of correct parse table.

    \section{The \parseit\ Tool}\label{sec:parseit}

\parseit\ takes as input a context free grammar and uses it as a
basis for generating questions. These questions are in the form
of MCQ~\footnote{MCQs have their advantages as well as
  disadvantages~\cite{mcq}. We chose MCQ as it is easier for the
  system to evaluate student choices as compared to the free
  form text answers.} and deal with various concepts related to
parsing.  The normal workflow involves the following steps:
\begin{enumerate}
\item The user  provides an  input grammar  and the  choice of  topic. The
  topics refer to  the concepts related to parsing such  as FIRST set,
  FOLLOW set, LL Parsing Table, LL Parsing Moves, LR(0) Item-sets, SLR
  Parsing Table, SLR Parsing Moves, etc.
\item A primary multiple choice question is generated based
  on the above two pieces of information.
\item If the user answers the problem incorrectly, then hints
  are generated for the same question in the form of
  questions.
\item When a correct solution to the problem is received,
  another question for the same topic is generated and
  presented to the user.
\end{enumerate}

In the  preprocessing step, the  system takes  a grammar as  input and
generates  the   information  required  for  correct   solutions.   In
particular, the  tool generates the FIRST  set and the FOLLOW  set for
all non-terminals,  LL Parsing  Table, LR(0)  items, canonical  set of
items for SLR parser, and SLR parsing table.

For primary problem for the selected topic, \parseit\ uses
the data-structures to form MCQs having multiple correct
answers.  Users have to select all valid options, and no
invalid option, for the answer to be deemed correct.
The options are also generated using the preprocessed data.

In the answer evaluation step, the solution given by the user
is compared  with the  solution computed by  the tool  in the
preprocessing step. If the  solutions match, then the control
transfers  back to  the  primary problem  generation step  to
generate  the  next question.  However,  if  the solution  is
wrong, the tool collects: a)  the incorrect options which are
selected, and  b) the correct  options which
are not  selected by  the user  and passes  them to  the hint
generation  step.\footnote{In the  rest of  the paper,  unless
  specified otherwise,  we use the term  incorrect choice for
  both the types of mistakes,  i.e., the missing valid choice
  and the selected invalid choice.}

For hints,  \parseit\  generates multiple hint  questions 
for each of  the incorrect choices.  These  questions are MCQs
having a single correct choice. These questions help the user to
revise the  concept required to  get correct solution  to the
primary question.

\subsection{Problem Generation}
Parsing techniques require solving three main types of problems: a)
computation of sets of elements, for example, FIRST, FOLLOW, LR Items,
GOTO, CLOSURE, b) computation of entries in a parse table, and
c) steps of a parser on a given input string. 

Since all the sets and tables are computed by \parseit\ in the
preprocessing step, generation of questions is easy.  The details
are given in a technical
report~\blind[0]{\cite{nimisha2015parsing}}. For
a question, the set of candidate choices is obtained by adding to
the set of correct choices a few mutations (addition/removal of a
term) or using a term from the solution for another similar
problem.  Evaluation of user solution is a simple comparison with
the computed solution.  

\subsection{Hint Generation} \label{subsec:hint generation}
\parseit\ generates 2 types of hint questions to guide  the user into
reaching the correct solution of the primary problem and understanding
her  mistakes.
\begin{itemize}
\item \textbf{\hintmcq:} Multiple choice questions are generated about
  the options selected by the user. These are typically generated when
  the user's solution for a primary problem either contains an
  incorrect option or omits a correct option\footnote{Hints are  sometimes generated  even when  the
  user  solution is  correct  (with a  small probability).  Otherwise,
  whenever the hint is generated, the user will know that her original
  answer  was  wrong, and  may  change  it without  understanding  the
  concept.}.  These help the user to
  understand that (a) the rules for the concept under test can not
  result in the particular incorrect option, and (b) the omitted
  option is a part of the correct solution and the rules that are used
  to get that option.
\item \textbf{\hintstring:} These types of questions are generated for
  filling LL/LR parsing tables of  grammars having no conflicts (i.e.,
  no duplicate  entries in the  cells).  If  the questioned cell  of a
  parsing  table is  filled incorrectly,  then the  tool automatically
  generates an input  string that exercises the contents  of that cell
  during parsing.  Due  to incorrectly filled entry,  the parsing with
  incorrect  table  for  the  string fails,  thus  hinting  user  that
  something is wrong with her solution.
\end{itemize}
\begin{example}  \label{ex:first-set} {\em
  Consider the grammar:
  \[ \begin{array}{rcl}
    E &\to &T + E \mid T \\
    T &\to &0 \mid 1\\
  \end{array}\]
  one of the questions generated for FIRST set is:\footnote{The interaction with \parseit\ (questions and hints generated) are formatted for readability.}
  \begin{description}
  \item{\bf Question:} Which symbols should be included in FIRST[T]?
  \item{\bf Options:}  (a) 1  (b) 0  (c) $+$ 
  \end{description}
  For a user selected symbol $\psi$, the hint questions are necessarily generated if the selection $\psi$ is incorrect. Even for a correct answer, the hints questions are generated with some low probability. The hint questions are of the form:	
  \begin{description}
  \item{\bf HintQ:} According to which of the following rules, the symbol $\psi$ is a part of FIRST[T]?
  \begin{enumerate}
  \item If $X$ is a terminal, then FIRST($X$) is $\{X\}$.
  \item If $X$ is a nonterminal and $X \to Y_1Y_2....Y_k$ is a
    production for some $k\geq 1$, then place $a$ in FIRST($X$) if for
    some $i$, $a \in $ FIRST($Y_i$), and $\epsilon$ is in all of\linebreak
    FIRST($Y_1$),\ldots,FIRST($Y_{i-1}$). If $\epsilon$ is in
    FIRST($Y_j$) for all $j = 1, 2, \ldots, k$, then add $\epsilon$
    to FIRST($X$).
  \item If $X\to\epsilon$ is a production, then add $\epsilon$
    to FIRST($X$).
  \item No valid rule for this symbol.
  \end{enumerate}
  \end{description}
  If this hint question is asked for incorrect option (`+'), the answer is option `4'. If a user selects any other option, the same question is repeated.
  For a correct option, say `0', the expected answer is option '2'.
  
  After these hint questions, question $Q_1$ is repeated. If it is answered correctly, then next question $Q_2$ is generated. Otherwise, the above process is repeated.
  }\mbox{}\hfill$\qed$
\end{example}

\begin{example}  \label{ex:ll_parse} {\em
    Consider the grammar:
  \[ \begin{array}{rcl}
    S &\to &a\ A\ B\ b \\
    A &\to &c \mid \epsilon\\
    B &\to &d \mid \epsilon\\
  \end{array}\]
  The following question (\hintmcq) is generated for the entries of LL parsing table:
  \begin{description}
  \item{\bf Question:} Which grammar rule should be included in the
    cell $[B, b]$ of the parsing table? (A partially filled parse
    table is shown to the user.)
  \item{\bf Options:} (a) $A \to c$ (b) $A \to \epsilon$ (c) $B \to d$
    (d) $B \to \epsilon$
  \end{description}
  For a production $\pi$ selected by the user, the hint questions generated could be:
  \begin{description}
  \item{\bf HintQ:} According to which of the following rules, the production $\pi$ is in cell $[B, b]$?
    \begin{enumerate}
    \item $\pi$ is production $B \to \alpha$, and $b \in \mbox{FIRST}(\alpha)$
    \item $\pi$ is production $B \to \alpha$, $\epsilon \in
      \mbox{FIRST}(\alpha)$ \\ and $b \in \mbox{FOLLOW}(B)$
    \end{enumerate}
  \end{description}
  }\mbox{}\hfill$\qed$
\end{example}

\begin{example}\label{ex:ll_parse_str} {\em
\newcommand{\exstr}{{\bf ab}}
For the question of Example~\ref{ex:ll_parse}, The correct answer is (d) $B \to \epsilon$. if a user selects an incorrect choice, \parseit\ can generate a string that will not be parsed correctly by the user's parse table. Say the user selects choice~(c) $B \to d$. \parseit\ will generate string \exstr, and ask the hint question (\hintstring):
\begin{description}
  \item{\bf HintQ:} LL-parsing on input \exstr\ with your parse table
    is failing.  Can you fix the error by selecting the correct
    choice? (Parser movements shown to the user omitted for brevity)
  \item{\bf Options:}  (a) $A \to c$ (b) $A \to \epsilon$ (c) $B \to \epsilon$
\end{description}
  }\mbox{}\hfill$\qed$
\end{example}

The most interesting feature of \parseit\ is the generation of an
input string to exercise incorrect entry in an incorrectly filled
parse table. We describe this in brief.

\section{Input String Generation}\label{sec:input}
A parsing  table is  a two-dimensional  array used  to parse  an input
string.  A grammar  is  considered LL/LR  if  the corresponding  LL/LR
parsing table has no duplicate entries.
We generate hints of type Hint\_String for grammars accepted by LL(0),
LL(1) or SLR parsers. For simplicity, we assume at most one mistake (in
some non-empty  entry) in the  table. If a  non-empty cell of  the parsing
table is  filled incorrectly by the  user, then there exists  a string
accepted by the grammar that exercises the cell, but can not be parsed
by  the user  filled (incorrect)  parsing table.   \parseit\ generates
such an input  string.  Then the corresponding question is  to ask the
user to  parse the string using  her incorrect table.  The  failure to
parse the string serves as a hint that the table has errors in it.

\newcommand{\gph}{\ensuremath{\mathcal{G}}}
\subsection{Input String Generation for LL Parsing}
\parseit\ creates  a graph \gph\ for  the grammar as  follows: All the
symbols in the grammar appear as  the vertices of the graph, edges are
created from the symbol on the left side to each of the symbols on the
right  side of  a rule.   An edge  is labeled  with the  shortest rule
(i.e. the  rule with the smallest number  of symbols on the  RHS) used for
creating  it.   For each  non-terminal,  \parseit\  also computes  and
records a terminal-only string derivable from that non-terminal. Terminal-only string for a terminal $t$ is $t$ itself.

Assume that, for  the incorrectly filled cell, $N$  represents the row
label (non-terminal)  and $t$ represents the  column label (terminal).
Let $S$ be  the start symbol of  the grammar.  We start  with a string
{\tt str1=$S$} and traverse \gph\ from start node (labeled $S$) to the
node  labeled  $N$  on  the shortest  path.   During  traversal,  the
non-terminal corresponding  to the  source vertex  in the  string {\tt
  str1} is replaced  by the label on the outgoing  edge traversed.  On
reaching the node labeled $N$, all symbols in {\tt str1}, except
last $N$, are replaced by their terminal-only strings.

In the next step, the tool  finds the shortest path from the node labeled
$N$ to the node labeled $t$.  This  path is traversed in the same
way as described above to give a string {\tt str2}.  $N$ in {\tt str1}
is replaced by {\tt str2} to  get the final input string. 

Note that  the process  described above is  similar to  constructing a
parse tree, the  difference being that we are creating  the tree for an
unknown string! The following example explains the algorithm.

\begin{figure}[t]
\centering
\begin{tabular}{@{}l@{}l@{}}
  \multicolumn{2}{@{}l@{}}{
    \renewcommand{\arraystretch}{1.1}
    \begin{tabular}[t]{|@{}l@{}|}\hline
      E $\to$ T \p{E} \\
      \p{E} $\to$ + T \p{E} | $\epsilon\;$ \\
      T $\to$ F \p{T} \\
      \p{T} $\to$ * F \p{T} | $\epsilon$ \\
      F $\to$ ( E ) | id \\ \hline
    \end{tabular}
    \begin{tabular}[t]{|@{}c@{}|@{}c@{}|@{}c@{}|@{}c@{}|@{}c@{}|@{}c@{}|@{}c@{}|} 
      \hline
      & \textbf{(} & \textbf{)} & \textbf{id} & \textbf{*} & \textbf{+} & \textbf{\$} \\
      \hline
      \textbf{T} & T$\to$F\p{T} &  & T$\to$F\p{T} &  &  &  \\
      \hline
      \textbf{F} & F$\to$(E) &  & F$\to$id &  &  &  \\
      \hline
      \textbf{E} & E$\to$T\p{E} &  & E$\to$T\p{E} &  &  &  \\
      \hline
      \textbf{\p{T}} &  & 
      \fcolorbox{lightgray}{lightgray}{\!\!\p{T}$\to\epsilon$\!\!} 
      &  & \ \p{T}$\to\;$ & \p{T}$\to\epsilon\;$ & \p{T}$\to\epsilon\;$ \\
      & &  \fcolorbox{lightgray}{lightgray}{\phantom{\!\!\p{T}$\to\epsilon$\!\!}}
      & & $*$F\p{T} & & \\
      \hline
      \textbf{\p{E}} &  & \p{E}$\to\epsilon$ &  &  & \p{E}$\to$ & \p{E}$\to\epsilon$ \\
                     &  &  &  &  & $+$T\p{E} &  \\
      \hline
    \end{tabular}
}
\\ \\
\multicolumn{2}{@{}l@{}}{
  (a) Grammar. \qquad (b) LL parsing table for the grammar.} \\ \\ \\
\multicolumn{2}{c}{
\includegraphics{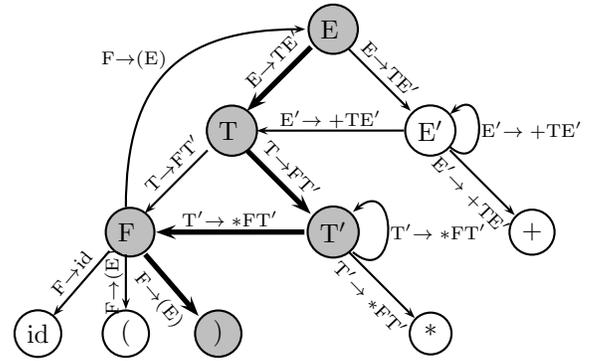}
}
\\ \\
\multicolumn{2}{l}{(c) Graph for the example grammar. The shortest paths}\\
\multicolumn{2}{l}{\phantom{(c)}  from $E$ to \p{T} and from \p{T} to $)$ are highlighted.}\\ \\
\end{tabular}
\caption{Input string generation for LL parsing.}
\label{fig:inp-for-ll}
\afterfig
\end{figure}

\begin{example}
\label{ex:LL input string}
{\em  Consider  the  grammar in  Fig.~\ref{fig:inp-for-ll}(a).   The
  correct  LL   parsing  table  M   for  this  grammar  is   shown  in
  Fig.~\ref{fig:inp-for-ll}(b). Fig.~\ref{fig:inp-for-ll}(c) shows  the
  graph generated by \parseit\ for this grammar.
  
  Suppose a user fills the cell M[\p{T}][)] incorrectly.  In order to
  build the desired input string to exercise this cell, \parseit\ will
  compute the shortest path from the start symbol E to \p{T} (the
  non-terminal representing the erroneous cell) using Dijkstra's
  shortest path algorithm (Fig.~\ref{fig:inp-for-ll}(c)). The
  traversal of this path results in the following sequence of strings
  being generated:
  \[ E \leadsto T\p{E} \leadsto F\p{T}\p{E} \]
  Since the target non-terminal, \p{T}, is reached, all other
  non-terminals are replaced by their terminal-only strings (id for
  F and $\epsilon$ for \p{E}):
  \[F\p{T}\p{E} \leadsto \mbox{id}\p{T}\epsilon \equiv \mbox{id}\p{T} \]
  Next \parseit\ finds the shortest path from \p{T} to )
  (Fig.~\ref{fig:inp-for-ll}(c)), and expands it:
  \[ \mbox{id}\p{T} \leadsto \mbox{id}*F\p{T} \leadsto
  \mbox{id}*(E)\p{T}\]  Non-terminals  are  replaced by  terminal-only
  strings (E  by id,  \p{T} by $\epsilon$)  to give the  desired input
  string:
  \[  \mbox{id}*(E)\p{T} \leadsto \mbox{id}*(\mbox{id}) \] 
  Note  that \p{T} is not treated especially again as we
  have already taken care of exercising the desired entry,
  M[\p{T}][)].}\hfill$\qed$
\end{example}

\newcommand{\state}{\ensuremath{s}}
\newcommand{\trans}{\ensuremath{\alpha}}
\newcommand{\stk}{\ensuremath{\mathcal{S}}}
\newcommand{\pinput}{{\em PStr}}
\newcommand{\nX}{{\em X}}
\newcommand{\prestr}{{\em Pref}}

\subsection{Input String Generation for LR Parsing}

The  parse table  for  LR  parsers consists  of  two sub-tables:  {\em
  Action}  table and  {\em Goto}  table~\cite{asu}.  The  action table
describes the  interaction between a  state in parsing and  a terminal
while the goto  table describes the interaction between a  state and a
non-terminal.  The action table has  two types of entries: {\em shift}
entries, that  push a  state on  the parsing  stack; and  {\em reduce}
entries that remove a certain number of states from the parsing stack.
The entries in goto table are  also shift entries (similar to those in
the action table) in that they also push a state in the parsing stack.

Both types of shift entries look only at the top of the parsing stack
and the {\em current} symbol (terminal or non-terminal) to decide the
move. For reduce entries, on the other hand, the top few entries of
the stack must {\em match} states corresponding to the right-hand side
of a rule.  As a result, the input string generation for LR parsing
needs to setup the parsing stack properly and uses heuristics to pick
appropriate reduce rules from the parse table to make sure that the
parser stack is in accept state at the end of input.

Assume L denotes an LR parsing table, \state\ represents the
row label (state) and \trans\ represents the column label
(non-terminal or terminal, including the special end marker
\$) for an incorrectly filled cell, L[\state][\trans].
\parseit\ computes a terminal-only string corresponding to
each non-terminal, and a deterministic finite automaton
(DFA) to recognize {\em viable prefixes}~\cite{asu}.  All
the states corresponding to the item-sets containing item(s)
with a dot at the end position are referred to as the {\em
  reduce} states of the DFA.  The tool maintains a stack, a
prefix string, and a set $X$ of possible next symbols.

Assume L denotes a LR parsing table, \state\ represents the row
label (state) and \trans\ represents the column label (non-
terminal or terminal, including the special end marker \$) for an
incorrectly filled cell, L[\state][\trans]. \parseit\ computes a
terminal-only string corresponding to each non-terminal, and a
deterministic finite automaton (DFA) to recognize {\em viable
  prefixes}~\cite{asu}. All the states corresponding to the
item-sets containing item(s) with dot at the end position are
referred to as the {\em reduce} states of the DFA. The tool
maintains a stack, a prefix string, and a set $X$ of possible
next symbols.
  
In the viable prefix DFA, a path from initial
state, $\state_0$ , to the state $\state$ is traversed. Labels for the nodes
(states) and the edges (terminal/non-terminals) along the path
are pushed on the stack in the order in which they appear during
traversal. The prefix string is generated by concatenating the
edge labels traversed, and replacing every non-terminal by its
terminal-only string. The next steps differs depending on
the type of correct entry in the cell L[\state][\trans].

\subsubsection{Shift Entry} 
If the correct entry for L[\state][\trans] is a shift entry (either in
action  or in  goto sub-table),  then,  in the  DFA, there  will be  an
outgoing edge with label \trans\ from the node for state $s$ to a state
$\state_\trans$.   Symbol   \trans\  is  pushed  onto  the   stack  and  its
terminal-only string  is appended to  the partial input  string. State
$\state_\trans$ is  also pushed on  the stack.  A  path is then  found from
$\state_\trans$ to  $\state_r$, one of the  reduce states\footnote{We chose
  the  shortest distance  reduce  state from  $\state_\trans$, to  generate
  shorter input  string.}.  All  the node labels  and the  edge labels
along the path  are pushed onto the stack in  that order.  The partial
input string is appended with the terminal-only string for the path.

The column labels (symbols)  corresponding to non-empty cells in row
for $\state_r$ in the parse table are added to the set of next symbol
$X$.  The next  action  is performed  on  the set  according to  the
following heuristics:
\begin{itemize}
\item  If row  $\state_r$ contains  both shift  and reduce  actions: We
  chose reduce action over shift actions\footnote{Note that all reduce
    actions in  a row will  be the same.}.  Any of the symbols  in $X$
  corresponding to the columns for  reduce action can come next in the
  input string  to force the reduction  on stack. We  differ the exact
  choice of symbol until we need to make a shift move (next point).
\item If row $\state_r$ contains only shift actions: We compute the
  number of symbols between the dot and the end of RHS in each rule
  for each state. We choose to shift to the state having the smallest
  value. In case of tie, any of the states involved in the tie is
  chosen. Partial string is updated by concatenating the terminal-only
  string corresponding to the column label of the entry chosen for
  shift.
\end{itemize}
The stack  is updated to  reflect the action chosen.   \parseit\ keeps
track of the  choice made at this step to avoid  using the same choice
repeatedly. This is necessary  to avoid generation of unbounded length
input. The process is repeated  until ``accept'' appears on top of the
stack. At this point, the partial input string is the desired
input. The following example illustrates these steps.

\begin{figure}[t]
\centering
\scalebox{.95}{
  \begin{tabular}[t]{@{}l@{}l@{}l@{}}
    \renewcommand{\arraystretch}{1.1}
    \begin{tabular}[t]{|@{}r@{\ }l@{}|}\hline
      0)& \p{S} $\to$ S \\
      1)& S $\to$ CC \\
      2)& C $\to$ aC \\
      3)& C $\to$ d \\\hline
      \multicolumn{2}{@{}l@{}}{(a) Augmented}\\
      \multicolumn{2}{@{}l@{}}{\phantom{(a)} Grammar} 
    \end{tabular}
    &
    \renewcommand{\arraystretch}{1.1}
    \begin{tabular}[t]{|@{}c@{}|@{}c@{}|}\hline
      \begin{tabular}[t]{@{}c@{}}        
        \begin{tabular}[t]{@{\ }r@{\ }p{11mm}}
          I0:& \p{S} $\to$ $\cdot$S \\
          & S $\to$ $\cdot$CC \\
          & C $\to$ $\cdot$aC \\
          & C $\to$ $\cdot$d \\
        \end{tabular} \\ \\
        \begin{tabular}{@{\ }r@{\ }p{11mm}}
          I1:& \p{S} $\to$ S$\cdot$
        \end{tabular}\\ \\
        \begin{tabular}{@{\ }r@{\ }p{11mm}}
          I2:& S $\to$ C$\cdot$C \\
          & C $\to$ $\cdot$aC \\
          & C $\to$ $\cdot$d \\
        \end{tabular}
      \end{tabular} &
      \begin{tabular}[t]{@{}c@{}}        
        \begin{tabular}[t]{@{\ }r@{\ }p{11mm}} 
          I3:& C $\to$ a$\cdot$C \\
          &     C $\to$ $\cdot$aC \\
          &    C $\to$ $\cdot$d \\
        \end{tabular}\\ \\
        \begin{tabular}{@{\ }r@{\ }p{11mm}}
          I4:& C $\to$ d$\cdot$ \\ 
        \end{tabular}\\ \\
        \begin{tabular}{@{\ }r@{\ }p{11mm}}
          I5:& S $\to$ CC$\cdot$ \\ 
        \end{tabular}\\ \\
        \begin{tabular}{@{\ }r@{\ }p{11mm}} 
          I6:& C $\to$ aC$\cdot$ \\
        \end{tabular} \\
      \end{tabular}\\ \hline
      \multicolumn{2}{c}{(b) Item sets}
    \end{tabular}
    &
    \renewcommand{\arraystretch}{1.1}
    \begin{tabular}[t]{|@{}c@{}||@{\ }c@{\ }|@{\ }c@{\ }|@{}c@{}||@{\ }c@{\ }|@{\ }c@{\ }|}
      \hline
      \textbf{S} & \multicolumn{3}{@{}c@{}}{\textbf{Action}\;\;\raisebox{-1.5mm}{\rule{.5pt}{12pt}}\,\raisebox{-1.5mm}{\rule{.5pt}{12pt}}} &
      \multicolumn{2}{@{}c@{}|}{\textbf{Goto}}\\
      \cline{2-6}
      \textbf{T} & \textbf{a} & \textbf{d} & \textbf{\$} & \textbf{S} & \textbf{C}\\
      \hline
      \textbf{0} & s3 & s4 &  & 1 & 2\\
      \hline
      \textbf{1} &  &  & acc &  & \\
      \hline
      \textbf{2} & s3 & s4 &  &  & 5\\
      \hline
      \textbf{3} & s3 & s4 &  &  & 6\\
      \hline
      \textbf{4} & r3 & r3 & r3 &  & \\
      \hline
      \textbf{5} &  &  & r1 &  & \\
      \hline
      \textbf{6} & r2 & r2 & r2 &  & \\
      \hline
      \multicolumn{6}{c}{(c) LR Parse Table} 
    \end{tabular}
  \end{tabular}} \\

  \begin{tabular}[t]{@{}c@{}}
    \includegraphics{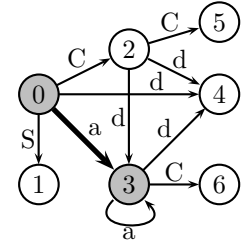}\\
    \scalebox{.95}{(d) DFA for viable prefixes. Path from state 0 to 3 is highlighted.}\\  \\
  \end{tabular}
  \caption{Input string generation for LR grammar.\label{fig:inp-lr-shift}}
\end{figure}

\begin{example}
  \label{ex:LR input string}  
  {\em
    Consider    the   augmented   grammar    in
    Fig.~\ref{fig:inp-lr-shift}(a).   The  item-sets,  the
    correct parse table, and the DFA for viable prefixes are
    shown in  Fig.~\ref{fig:inp-lr-shift}(b), (c)  and (d)
    respectively.  For the DFA, state 0 is the initial state
    and \{1, 4, 5, 6\} is the set of reduce states.
    
    Suppose user makes an incorrect entry in the cell
    L[3][d], then the shortest path from state 0 to state 3
    is traversed (Fig.~\ref{fig:inp-lr-shift}(d)).  From
    state 3, there is a transition on symbol d to state 4,
    which happens to a reduce state.  The stack (\stk),
    partial string (\pinput), and set of potential next
    symbols (\nX) maintained by \parseit\ at this stage are
    shown in Fig.~\ref{fig:incorrect-shift-lr}(a). 
    
    The non-empty entries in the parse table for the row of
    state 4 (top of the stack) are considered for the next
    set of the terminal symbols. For this table, all entries
    are same (r3), so all of them are considered for
    application of reduced rule.
    Fig.~\ref{fig:incorrect-shift-lr}(b) shows the
    configuration.

     We repeat the process with state 6 on the top of the
     stack. In this case, the same columns have non empty
     entries (r2). So, there is no refinement in the choice
     of the next symbol, and the set \nX\ remains the
     same. The reduction results in the
     configuration Fig.~\ref{fig:incorrect-shift-lr}(c).

     Now state 2 is on top of the stack. The corresponding
     row contains non empty entries, s3 and s4, for columns
     labeled a and d. This restricts the choice for next
     symbol to \{a, d\} (Fig.~\ref{fig:incorrect-shift-lr}(d)).

     Using the heuristic to shift to a state where dot is
     closest to the right end of RHS of some rule, we prefer
     a shift to state 4 (C$\to$d$\cdot$ has $\cdot$ at the
     right end, compared to C $\to$ a$\cdot$C in state 3,
     where $\cdot$ is 1 symbol away from the right end.) This
     forces the next symbol to be d. Since we have narrowed
     down the next symbol to a unique choice, the symbol is
     appended to the partial string and corresponding shift
     is made on the stack. Set \nX\ is cleared
     (Fig.~\ref{fig:incorrect-shift-lr}(e)).

     The process is continued till we get an ``accept'' on the
     stack. The remaining configurations for the example are
     shown in Fig.~\ref{fig:incorrect-shift-lr}(f).
     The desired input string to exercise L[3][d]  is  partial string at this
     stage (ignoring the end-marker \$), i.e. {\tt add}.}
\mbox{}\hfill$\qed$
\end{example}

\begin{figure}[t]
  \centering
  \begin{tabular}{@{}c@{\ \ }c@{}}
    \begin{tabular}[b]{|@{\ }l@{\ }|@{\ }l@{\ }|@{\ }l@{\ }|}
      \hline
      \textbf{\stk} & \textbf{\pinput} & \textbf{\nX}\\
      \hline
      0a3d4 & ad & \\
      \hline
    \end{tabular}
    &
    \begin{tabular}[b]{|@{\ }l@{\ }|@{\ }l@{\ }|@{\ }l@{\ }|}
      \hline
      \textbf{\stk} & \textbf{\pinput} & \textbf{\nX}\\
      \hline
      0a3d4 & ad & \{a, d, \$\} \\
      0a3C6 & ad & \\
      \hline
    \end{tabular}
    \\
    (a) & (b) \\  & \\
    \begin{tabular}[b]{ |@{\ }l@{\ }|@{\ }l@{\ }|@{\ }l@{\ }|}
      \hline
      \textbf{\stk} & \textbf{\pinput} & \textbf{\nX} \\
      \hline
      0a3d4 & ad & \{a, d, \$\}\\
      0a3C6 & ad & \{a, d, \$\}\\
      0C2 & ad & \\
      \hline
    \end{tabular}
    &
    \begin{tabular}[b]{ |@{\ }l@{\ }|@{\ }l@{\ }|@{\ }l@{\ }|}
      \hline
      \textbf{\stk} & \textbf{\pinput} & \textbf{\nX} \\
      \hline
      0a3d4 & ad & \{a, d, \$\}\\
      0a3C6 & ad & \{a, d, \$\}\\
      0C2 & ad & \{a, d\} \\
      \hline
    \end{tabular}
    \\
    (c) & (d) \\  & \\
    \begin{tabular}{|@{\ }l@{\ }|@{\ }l@{\ }|@{\ }l@{\ }|}
      \hline
      \textbf{\stk} & \textbf{\pinput} & \textbf{\nX} \\
      \hline
      0a3d4 & ad & \{a, d, \$\}\\
      0a3C6 & ad & \{a, d, \$\}\\
      0C2 & ad & \{a, d\}\\
      0C2d4 & add & \\
      \hline
    \end{tabular}
    &
    \begin{tabular}{|@{\ }l@{\ }|@{\ }l@{\ }|@{\ }l@{\ }|}
      \hline
      \textbf{\stk} & \textbf{\pinput} & \textbf{\nX} \\
      \hline
      0C2d4& add & \{a, d, \$\}\\
      0C2C5 & add & \{\$\}\\
      0S1 & add\$ & \\
      accept & add\$ &  \\
      \hline
    \end{tabular}\\ 
    (e) & (f) \\
  \end{tabular}
  \caption{Configurations of \parseit\ while generating
    string for incorrect shift entry for LR
    grammar.\label{fig:incorrect-shift-lr}}
\end{figure}

\paragraph{Reduce Entry}
For generating input corresponding to a shift entry in a cell
in the  parse table  (more precisely, in  the goto  or action
sub-table),  the state  only at  the top  of parsing  stack is
important. For a reduce entry,  however, the  configuration of
stack below the  top also plays a vital role as  the RHS of the
rule,  using  which reduction  takes  place,  must match  the
prefix  at  the top  of  the  stack. Thus,  input  generation
algorithm not only needs to construct the next symbols in the
(unknown)  string, but  also the  previous symbols  that have
been already pushed on the stack (and possibly undergone some
reductions as well).

For  an incorrect  entry corresponding  to  the reduce  entry in  cell
L[\state][\trans], \parseit\ generates the input for reduce entries in
two  stages: in  the first  stage, the  stack is  setup to  enable the
application  of the  desired  reduce rule,  i.e.,  the symbols  before
\trans\  in  the input  string  are  guessed.   In the  second  stage,
\parseit\ generates the  symbols that may follow \trans\  in the input
string, till  the end (\$).  Note  that, \trans\ can itself  be \$, in
which  case  no  next  symbol   is  needed.  The  second  stage  works
identically  to the  generation of  input string  for incorrect  shift
entry.

Suppose  the correct  reduce entry  in the  cell L[\state][\trans]  is
$N\to\sigma$,    then   \parseit\    guesses   the    stack   to    be
$\Omega\state_b\Sigma\state$,  where  $\Omega\state_b$ represents  the
sequence  of states  corresponding to an unknown prefix  of the  desired
input,  and  $\Sigma$  is  the sequence  of  states  corresponding  to
$\sigma$, the RHS of correct reduce rule in the cell of interest.  The
input character  now seen by the  parser is \trans\ while  a prefix
string saves  the already  seen string--the terminal  string derivable
from symbols in  $\sigma$.  After reduction by  the rule $N\to\sigma$,
the  stack  configuration  becomes $\Omega\state_b  N\state_a$,  where
$\state_a$    is   obtained    from   the    goto   sub-table    entry
L[$\state_b$][$N$].   

If $\state_b$ is known, we can  find a satisfying prefix $\omega$ (the
stack configuration $\Omega$) by traversing the viable prefix DFA from
start  state to  $\state_b$, and  generating the  terminal-only string
corresponding  to the  traversed  edge labels  (node  labels and  edge
labels for $\Omega$).  However, since we do not know  these states, we
use  non-empty  entries in  the  parse  table  to guess  the  possible
choices:
\begin{eqnarray*}
  S_B &=& \{ \state_b \mid L[\state_b][N] \mbox{ is non-empty} \} \\
  S_A &=& \{ \state_a \mid L[\state_b][N] \mbox{ is }\state_a \}
\end{eqnarray*}
Thus,  $S_B$ represents  the set  of possible  states before  $N$, and
$S_A$  represents  the  set  of  possible  states  after  $N$  on  the
stack. Suppose  $t$ is  the input  character seen  by the  parser.  We
refine the choices using the following rules:
\begin{enumerate}
\item For $\state_a \in S_A$,  if L[$\state_a$][$t$] is error (empty),
  then $\state_a$  is removed from  $S_A$, and all $\state_b  \in S_B$
  such that  $ L[\state_b][N] \mbox{  is } \state_a$ are  removed from
  $S_B$.
\item If there is a state $\state_a  \in S_A$ which has shift entry in
  L[$\state_a$][$t$], we choose  it, and ignore all  other states from
  $S_A$. One of the corresponding states in $S_B$ (i.e., $\state_b \in
  S_B$ such that $L[\state_b][N] \mbox{  is } \state_a$) is chosen for
  the before state.  We now  have a shift entry, L[$\state_a$][$t$] to
  work on, which can be solved as described earlier.
\item If there is  no shift entry, we are left  with the reduce entries. We
  choose  any such  state  $\state_a  \in S_A$.   We  now have  another
  instance of  reduce entry  problem L[$\state_a$][$t$], which  can be
  solved  recursively.  If  there is  a unit-production  cycle in  the
  grammar,  care   must  be  taken   to  avoid  choosing   the  cyclic
  productions. This can be achieved by marking the productions already
  used.
\end{enumerate}
The working of \parseit\ for reduce entry problem is described in
the following example:
\begin{figure}[t!]
\centering
  \begin{tabular}{@{}c@{\ \ }c@{}}
    \begin{tabular}[t]{@{}c@{}}
      \begin{tabular}{|l|l|l|} 
        \hline
        \textbf{\stk} & \textbf{\pinput} & \textbf{\prestr}\\
        \hline
        $\Omega\state_b$d4 & a & $\omega$d\\
        \hline
      \end{tabular} \\ (a) \\ \\
      \begin{tabular}{|l|l|l|}
        \hline
        \textbf{\stk} & \textbf{\pinput} & \textbf{\prestr}\\
	\hline
	$\Omega$0C2 & a & $\omega$d \\
	\hline
      \end{tabular} \\ (c) \\ \\
      \begin{tabular}{|l|l|l|}
        \hline
        \textbf{\stk} & \textbf{\pinput} & \textbf{\prestr}\\
	\hline
	0C2 & a & d \\
	\hline
      \end{tabular} \\ (d)
    \end{tabular}
    &
    \begin{tabular}[t]{@{}c@{}}
      \begin{tabular}{|l|l|l|}
	\hline
	\textbf{\stk} & \textbf{\pinput} & \textbf{\prestr}\\
	\hline
	$\Omega\state_b$C$\state_a$ & a & $\omega$d \\
	\hline
      \end{tabular} \\ (b) \\ \\
      \begin{tabular}{|l|l|l|}
        \hline
        \textbf{\stk} & \textbf{\pinput} & \textbf{\nX}\\
        \hline
        0C2a3 & da & \{a, d\} \\
        0C2a3d4 & dad & \{a, d, \$\} \\
        0C2a3C6 & dad & \{a, d, \$\} \\
        0C2C5 & dad & \{\$\} \\
        0S1 & dad\$ &  \\
        accept & dad\$ & \\
        \hline
      \end{tabular} \\      
      (e)
    \end{tabular}
  \end{tabular}
  \caption{Configurations of \parseit\ while generating string
    for incorrect reduce entry for LR grammar.\label{fig:lr-reduce-ex}}
\end{figure}

\begin{figure*}[t!]
  \centering
  \includegraphics[width=1.7\columnwidth]{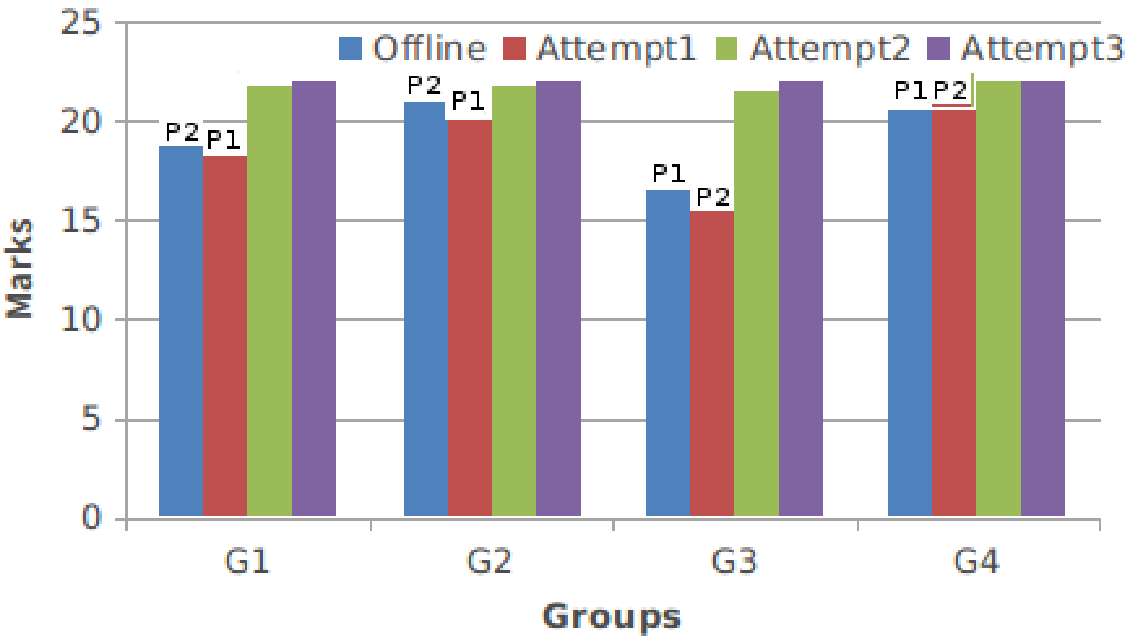}\\
 {Attempt2 (Attempt3) means correct solutions using
   up to one (two) hint(s) from \parseit.}
 \caption{Comparing group averages: offline vs. \parseit.\label{fig:avg}}
\end{figure*}
\begin{figure*}[t!]
\centering
\includegraphics[width=1.7\columnwidth]{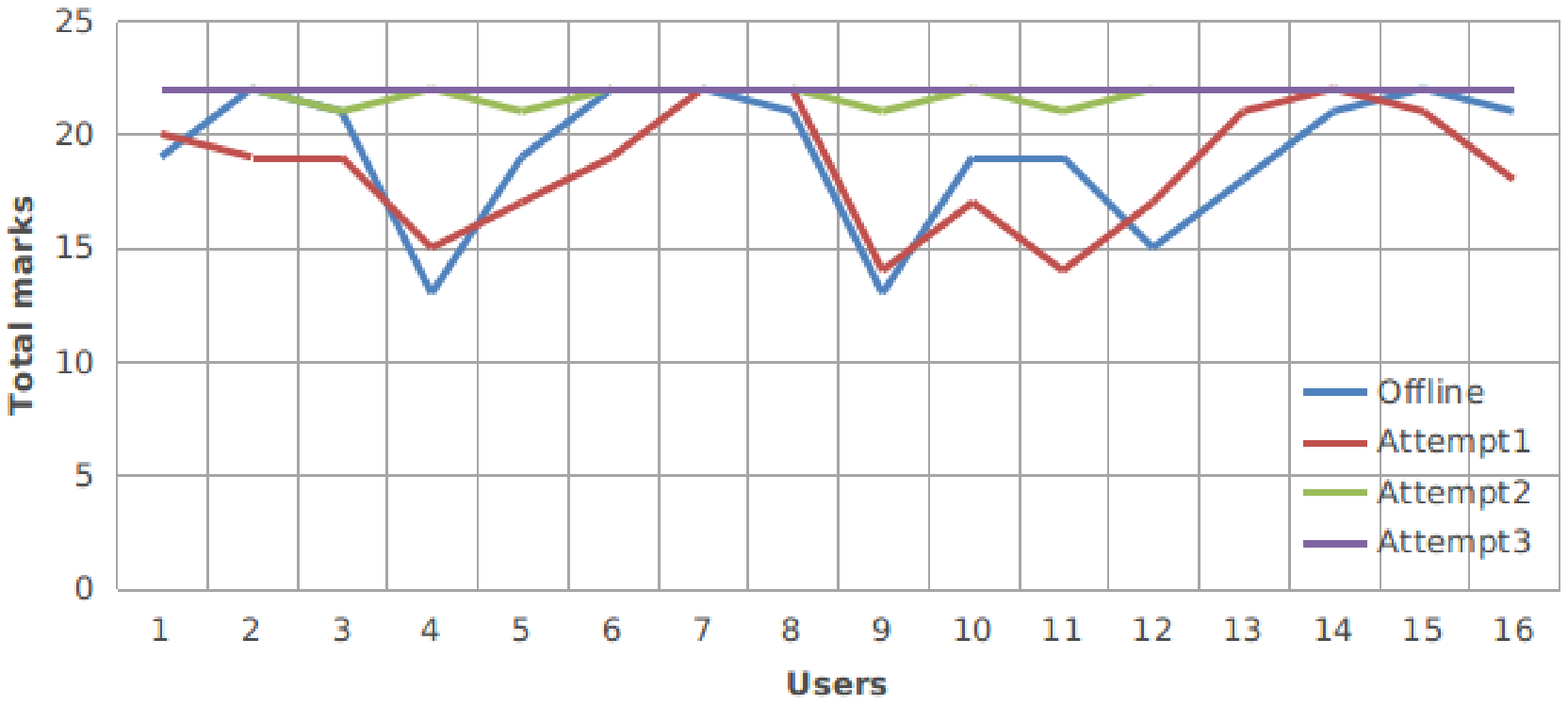}\\
 {Attempt2 (Attempt3) means correct solutions using
   up to one (two) hint(s) from \parseit.}
 \caption{Comparing individual averages: offline vs. \parseit.\label{fig:indiv}}
\afterfig
\end{figure*}

\begin{example}\label{ex:lr-reduce}
  {\em Consider the grammar and parse table in
    Fig.~\ref{fig:inp-lr-shift}. Suppose user makes an
    incorrect entry in the cell L[$4$][$a$] which contains entry
    ``r3'' i.e reduce by $C \to d$. The stack (\stk), partial
    string (\pinput), and the prefix string(\prestr)
    configuration maintained by \parseit\ at this stage are shown
    in Fig.~\ref{fig:lr-reduce-ex}(a). Here, d is the RHS of the
    reduce rule, as well as the terminal-only string.  Applying
    the reduce rule, gives the configuration in Fig~\ref{fig:lr-reduce-ex}(b).
    
    From parse table, the possible
    choices for  $\state_b$ are $S_B =  \{0, 2, 3\}$, and  the choices
    for $\state_a$ are $S_A = \{2, 5, 6\}$. We refine the choices using the steps defined above:
    \begin{enumerate}
    \item Since  L[5][a] is an  error entry,  state 5 is  removed from
      $S_A$ and corresponding state 2 is removed from $S_B$.
    \item Out of the remaining states in $S_A$, L[2][a]  has a shift
      entry and L[6][a] has a reduce entry. We chose 2 for $\state_a$,
      and corresponding state 0 for $\state_b$.
    \end{enumerate}
    We get configuration in Fig~\ref{fig:lr-reduce-ex}(c).  Since
    $\state_b$ is 0, the start state itself, $\omega$ is empty
    string (Fig~\ref{fig:lr-reduce-ex}(d)).

    The problem is now changed to shift entry problem L[2][a],
    having partial string {\tt da} (prefix and partial string
    concatenated). The configurations during the solution for the
    shift problem are as in Fig~\ref{fig:lr-reduce-ex}(e).
    So, {\tt dad} is the desired input to exercise cell L[4][a].
  }\mbox{}\hfill$\qed$
\end{example}

    \section{User Study}\label{sec:study}
To verify the effectiveness of \parseit, we
implemented the tool in Java. The prototype implementation is
available as a JAR file from anonymous Dropbox
link~\cite{parseitcode}. A web interface was created for the user
study.

The user study was conducted with with 16 students who have
already done an introductory course on {\em Compiler Design}. We
used 2 grammars and created 22 questions of 1 mark each related
to various sub-topics in parsing. The question papers are code
named P1 and P2. The students were randomly divided into 4 groups
of 4 students each, G1--G4.

Each group  solved one  question paper using  \parseit, and  the other
using pen and paper (offline mode). To maintain equality between the
two  approaches, we  provided a  cheat sheet  containing the  required
rules to  each student  for offline mode.  Further, the  sequence of
\parseit\ mode and  offline mode was alternated.  In particular, the
groups solved the grammar in the following order:
\begin{center}
  \begin{tabular}{|@{}r@{:\ }l@{}|} \hline
    G1&  P2 using offline followed by P1 using \parseit\\
    G2&  P1 using \parseit\ followed by P2 using offline\\
    G3&  P2 using \parseit\ followed by P1 using offline\\
    G4&  P1 using offline followed by P2 using \parseit\\\hline
  \end{tabular}
\end{center}

The students were asked to fill a survey about the effectiveness
of \parseit\ after solving both the papers.

  Fig.~\ref{fig:avg}
shows the average of marks for groups while Fig.~\ref{fig:indiv}
shows average marks for individuals with and without \parseit.
Comparing the average marks across sessions, we found that
average marks for G4 remain unchanged while for G1 it reduced by
0.5.  For both G2 and G3, the average marks increased by 1. If we
include the correct answer after a hint is taken during \parseit\
mode, we found that most student could get nearly full marks
across the groups (average over all students improved to 21.75
from 18.50 for \parseit\ without hints and 19.18 for offline).
The biggest improvement was of 7 marks, for 3 students.

Even though the data set is  small, it shows that online platform
itself does  not make  a big difference  in the  understanding of
parser  concepts,  but  the  hints'  mechanism  that  results  in
improvements in marks. The hints  allow students to correct their
mistakes  early. It  is also  easy to  figure out  the source  of
confusion for students,  which can be of help  to the instructor.
The  post  study  survey  also  corroborated  our  inference:  15
students accepted that the  hints provided a better understanding
of parsing, and  helped reach the correct  solution.  One student
had a negative feedback as he commented that ``Hints are produced
as questions, which  increases confusion as the  user has already
answered it  wrong.''.  However, generating hints  in other forms
(say natural  language sentence) is  an area of  future long-term
research. 
    
    \section{Conclusions and Future Work}\label{sec:concl}
In this paper, we described \parseit\ for teaching parsing techniques.
Our  approach  is  question-answering based:  problems  are  generated
automatically  and given  to  students  to explain  the  working of  a
parser. Further, the  hints provided by the tool are  also in forms of
targeted questions that help a student discover her mistake and revise
the concept at  the same time. \parseit\ allows students  to learn the
techniques at  their own  pace, according  to their  convenience.  The
user study shows  that the interactive nature of \parseit\ helps users
to learn from  their own mistakes through experiments,  and reduce the
burden on teachers and teaching assistants. 

Similar tools exist to teach few other phases of a compiler.  In
future, we plan to integrate these tools with \parseit, and
develop new tools to automate tutoring of all the phases of the
compiler.  We also plan to build animations around these concepts
to improve student experience and understanding.  An interesting
question that will require user study over a longer period is
whether the hints just helps the students select the correct
answer during the exam, or do they have a lasting learning
effect. Our plan is to deploy \parseit\ in a large class teaching
Compilers, to understand its impact on learning.

    \balance
    \bibliographystyle{abbrv}
    \bibliography{references}
\end{document}